\newcommand{\be}{\begin{equation}}
\newcommand{\ee}{\end{equation}}
\documentclass[twocolumn,showpacs,superscriptaddress,amsmath,amssymb]{revtex4}
\usepackage{graphicx}
\usepackage{epstopdf}
\usepackage{dcolumn}
\usepackage{bm}

\begin{document}
\title{Variational approach vs accessible soliton approximation in
nonlocal nonlinear media\\}

\author{Branislav N. Aleksi\'c}
\affiliation{Institute of Physics, University of Belgrade, P.O.Box
68, 11080 Belgrade, Serbia} \affiliation{Texas A\&M University at
Qatar, P.O.Box 23874, Doha, Qatar}

\author{Najdan B. Aleksi\'c}
\affiliation{Institute of Physics, University of Belgrade, P.O.Box
68, 11080 Belgrade, Serbia} \affiliation{Texas A\&M University at
Qatar, P.O.Box 23874, Doha, Qatar}

\author{Milan S. Petrovi\'c}
\affiliation{Texas A\&M University at Qatar, P.O.Box 23874, Doha,
Qatar} \affiliation{Institute of Physics, P.O.Box 57, 11001
Belgrade, Serbia}

\author{Aleksandra I. Strini\'c}
\affiliation{Institute of Physics, University of Belgrade, P.O.Box
68, 11080 Belgrade, Serbia} \affiliation{Texas A\&M University at
Qatar, P.O.Box 23874, Doha, Qatar}

\author{Milivoj R. Beli\'c}
\affiliation{Texas A\&M University at Qatar, P.O.Box 23874, Doha,
Qatar}


\begin{abstract}
We discuss differences between the variational approach to solitons and the accessible
soliton approximaion in a highly nonlocal nonlinear medium. We compare results
of both approximations by considering the same system of equations
in the same spatial region, under the same boundary conditions.
We also compare these approximations with the numerical solution of the equations.
We find that the variational highly nonlocal approximation provides more accurate
results and as such is more appropriate solution than the accessible soliton approximation.
The accessible soliton model offers a radical simplification
in the treatment of highly nonlocal nonlinear media, with easy comprehension and
convenient parallels to quantum harmonic oscillator,
however with a hefty price tag: a
systematic numerical discrepancy of up to 100\% with the numerical results.
\end{abstract}


\maketitle
\section{Introduction}

Self-localized wave packets which propagate in a nonlinear medium
without changing their structure are known as the optical spatial
solitons \cite{yuri}. Their existence is a consequence of the robust
balance between dispersion and nonlinearity or between
diffraction and nonlinearity or between the all three in the
propagation of spatiotemporal solitons or light bullets.
An important characteristic of many nonlinear media is their nonlocality,
i.e. the fact that the characteristic size of
the response of the medium is wider than the size of the excitation
itself. Strong
nonlocality is of special interest, because it is observed
in many media. For example, in nematic liquid crystals (NLCs) both
experimental and theoretical studies demonstrated that the
nonlinearity is highly nonlocal \cite{hen,cyril,beec}.

In 1997, Snyder and Mitchell introduced a model of nonlinearity
whose response is highly nonlocal \cite{snyder} -- in fact,
infinitely nonlocal. They proposed an
elegant theoretical model, intimately connected with the linear
harmonic oscillator, which describes complex soliton-like dynamics
(collisions, interactions, and deformations) in simple terms, even in two and
three dimensions. Because of the simplicity of the theory, they
coined the term "accessible solitons" (AS) for these optical
spatial solitary waves. But straightforward application of the AS
theory, even in nonlinear media with almost infinite range
of nonlocality, inevitably led to additional problems
\cite{assanto 2003,assanto 2004,henninot}, because there exists no
physical medium without boundaries and without noise.

To include interactions between solitons within
boundaries, as well the impact of the finite size of the sample,
we developed a variational approach (VA) to solitons in
nonlinear media  with long-range nonlocality, such as NLCs \cite{aleksic}.
Starting from a convenient ansatz, this approach delivers a stationary solution
for the beam amplitude and width, as well as the period of small
oscillations about the stationary state. It provides for natural explanation
of oscillations seen when, e.g. noise is included into the nonlocal nonlinear models.
The noise is inevitable in any real physical system and causes a
regular oscillation of soliton parameters with the period well predicted
by our VA calculus \cite{aleksic}. It may even destroy solitons \cite{petrovic}. Even though our
VA results were corroborated by numerics and experiments, they still attracted
a volatile exchange with other researchers \cite{comment,reply}.
We have further investigated the destructive
influence of noise on the shape-invariant solitons in a highly nonlocal
NLCs in \cite{petrovic}.

In this study of VA and AS approximations to the fundamental
soliton solutions in a (2+1)-dimensional highly nonlocal medium, we
adopt the following model of coupled normalized equations \cite{yuri,aleksic}:

\be
2i{\frac{\partial E}{\partial z}}+\Delta E+\vartheta E=0,
\label{eq1} \ee

\be
2\Delta \vartheta +\left\vert E\right\vert ^{2}=\alpha\vartheta, \label{eq2}
\ee

\noindent with zero boundary conditions on the border of a square
transverse region $\left\vert x\right\vert \leq d$ and $\left\vert
y\right\vert \leq d.$ Here $z$ is the propagation direction and $\Delta$
the transverse Laplacian. The system of equations of interest consists
of the nonlinear Schr\"{o}dinger equation for the propagation of the
optical field $E$ and the diffusion equation for the nonlocal response
of the medium $\vartheta$. This is a fairly general model for the
nonlinear optical media with a diffusive nonlocality, widely used in
the literature \cite{yuri,assanto 2003,aleksic}. In the local limit,
the first term in Eq. (\ref{eq2}) can be neglected and the model
reduces to the Schr\"{o}dinger equation with the Kerr nonlinearity.
In the opposite limit, the third term in
Eq. (\ref{eq2}) can be neglected and the highly nonlocal model
is reached. Since we are interested in the strong nonlocality,
we will omit in our analysis the term on the right-hand side of Eq. (\ref{eq2}).

\section{Variational Approach}

In this approach, to derive equations describing
evolution of an approximate field beam, a Lagrangian density is introduced,
corresponding to equations (\ref{eq1}, \ref{eq2}):

\be
\mathcal{L}=i\left( {\frac{\partial E^{\ast }}{\partial z}}%
E-{\frac{\partial E}{\partial z}}E^{\ast }\right) +\left\vert
\nabla E\right\vert ^{2}+\left\vert \nabla \vartheta \right\vert
^{2}-\vartheta \left\vert E\right\vert ^{2}.\qquad \label{lagrang}
\ee

\noindent Thus, the problem is reformulated into a variational problem

\be \delta \iiint \mathcal{L}dxdydz=0, \ee

\noindent whose solution
is equivalent to equations (\ref{eq1}, \ref{eq2}). To obtain
evolution equations for an approximate field in the highly nonlocal
region, an ansatz is introduced
in the form of a Gaussian beam for the field
\cite{aleksic}:

\be
E=A\exp [-{\frac{r^2}{2 R^2}}+iCr^2+i\psi ] , \label{trial1}
\ee

\noindent in which $A$ is the amplitude, $R$ is the beam width, $C$ is
the wave front curvature along the transverse coordinate, and $\psi$
is the phase shift. Variational optimization
of these beam parameters will lead to the most appropriate VA solution of the problem.
Likewise, a trial function
for the nonlocal response of the medium is introduced, in the form

\be
\vartheta =B\ \left[   \textrm{Ei}(-r^{2}/T^{2})-\ln
(r^2/d^2)\right] , \label{trial2}
\ee

\noindent which is characterized by the amplitude $B$ and the width $T$.
Here, Ei is the exponential integral function. Note that $T$ does not represent the
total width of $\vartheta$.
The form of $\vartheta$ corresponds to a radially-symmetric
solution of Eq. (\ref{eq2}), with zero boundary
conditions on a circle of radius $d \gg R$ (the limit of a thick
cell) \cite{abramowitz}. We take this expression as an approximate solution
on a square sample.

Let $\delta =\max {R^2/d^2} \ll 1$; then the
averaged Lagrangian $L=\iint \mathcal{L}dxdy$ is given by:

\be
\begin{split}
L &=2P\psi ^{\prime }+2PR^{2}\left( C^{\prime }+2C^{2}\right)
+\frac{P}{R^{2}} +\\
&+ 4\pi B^{2}\ln \left( \frac{e^{\gamma }d^{2}}{2T^{2}}\right)
-PB\ln \left( \frac{e^{\gamma }d^{2}}{R^{2}+T^{2}}\right)
+O(\delta ).
\end{split}
\label{lagr} \ee where $\gamma$ is Euler's constant and the prime
denotes the derivative with respect to $z$.
In the process of optimization from the averaged
Lagrangian, one obtains four ODEs:

\be
\frac{dP}{dz}=0 , \label{eul1}
\ee

\be
C=\frac{1}{2R}\frac{dR}{dz} , \label{eul2}
\ee

\be
\frac{d^2R}{dz^2}=\frac{1}{R^3}-\frac{P}{16\pi R}+O(\delta ) ,
\label{eul3}
\ee

\be
\frac{d\psi }{dz}=-\frac{1}{R^2}+\frac{P}{16\pi }\ln \left( \frac{
e^{\gamma +1/2}d^2}{2R^2}\right) +O(\delta ) , \label{eul4}
\ee

\noindent and two algebraic relations $T=R+O(\delta )$ and
$B=P/8\pi +O(\delta )$. The beam power $P=\int d\varphi \int |E|^2
rdr=\pi A^2 R^2$ is conserved, according to Eq.
(\ref{eul1}). The system of equations (\ref{eul1}-\ref{eul3})
describes the dynamics of the beam around a stationary state.

In the stationary state $(dR/dz=dC/dz=0)$, we find the equilibrium
beam width $R_0$ as a function of the beam power only:

\be
R_{0VA}=4\sqrt{\frac{\pi }{P}}+O(\delta ) . \label{stac1}
\ee

\noindent From relations (\ref{trial2}, \ref{eul4}) we also find
the maximum value of $\vartheta$:

\be
\vartheta _{\max (0)}=\frac{P}{8\pi }\ln \left(
\frac{e^{\gamma }}{16\pi } d^2 P^2 \right) +O(\delta ) ,
\label{stac2}
\ee

\noindent and the propagation constant $\mu =\left( d\psi /dz\right)
_{0}$ can be written as:

\be
\mu =\frac{P}{16\pi }\ln \left( \frac{e^{\gamma -1/2}}{32\pi }
d^2 P^2 \right) +O(\delta ) .  \label{stac3}
\ee

\noindent It should perhaps be mentioned that the integral quantity $\Theta =2\pi
\int_{0}^{d}\vartheta rdr=P\left( d^2-R^2\right) /8+O(\delta )$
$\approx P d^2/8$, which is proportional to the power, is also
conserved.

The period of small oscillations of the perturbation around the
equilibrium position ($R=R_0$, $C=C_0$) is given by the following
relation:

\be
\Lambda_{VA}=\frac{16\sqrt{2} \pi^2}{P} . \label{period VA}
\ee Relations (\ref{stac1} - \ref{period VA}) completely define
the VA approximate solution in the highly nonlocal case. It
remains to do the same for the AS approximation and then to
compare the two.

\section{Accessible soliton approximation}

In the AS approximation, the basic assumptions
are that the shape of the nonlocal response of the medium is
a parabolic function of the transverse distance,

\be
\vartheta =\theta _{0}\ -\theta _{2}r^2 ,  \label{pot}
\ee

\noindent and that the shape function of the field $E$ is still a Gaussian,
given by equation
(\ref{trial1}). The only refractive index "seen" by the beam is
that confined near its propagation axis \cite{snyder}.

The parameters of the trial function (\ref{trial1}) are now given by equations:

\be
\frac{dA}{dz}=2AC ,
\ee

\be
C=\frac{1}{2R}\frac{dR}{dz} , \label{as1}
\ee

\be
\frac{d^2 R}{dz^2}=\frac{1}{R^3}-\theta _2 R , \label{as2}
\ee

\be
\frac{d\psi }{dz}=-\frac{1}{R^2}+\frac{\theta _0}{2} ,
\label{as3}
\ee

\noindent and equation (\ref{pot}), which exactly satisfy equation (\ref{eq1}).

\begin{figure}\vspace{0mm}
\includegraphics[width=70mm]{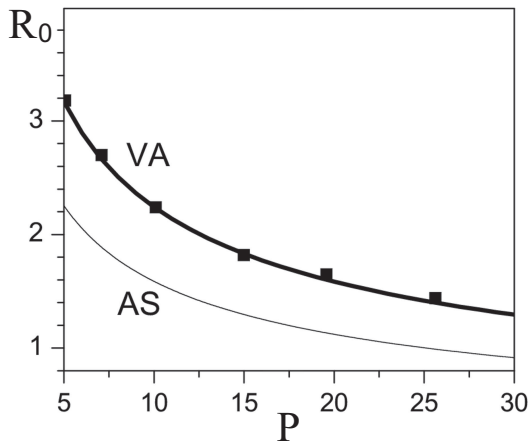}\vspace{0mm}
\caption{\label{Fig1}
Stationary beam width as a function
of the beam power, in both approximations.
Dots represent results obtained in numerical simulations.
The box width is $d=40$.}
\vspace{0mm}
\end{figure}

The parameter $\theta_0$ is only a phase shift and as such quite
arbitrary.
On the other hand, the value of $\theta_2$ is much more important;
it is determined from equation
(\ref{eq2}). By replacing (\ref{trial1}) and (\ref{pot}) in
equation (\ref{eq2}), in the limit $r\rightarrow 0$ one obtains
$\theta _2=A^2/8$ \cite{assanto 2004}. Then, equation
(\ref{as2}) becomes:

\be
\frac{d^ 2R}{dz^2}=\frac{1}{R^3}-\frac{P}{8\pi R} .
\label{as4}
\ee

\begin{figure}\vspace{0mm}
\includegraphics[width=70mm]{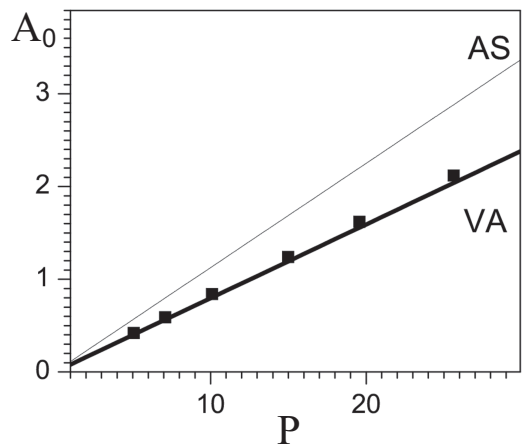}\vspace{0mm}
\caption{\label{Fig2} Stationary amplitude as a function of the
beam power, in both approximations.
} \vspace{0mm}
\end{figure}

\noindent The equilibrium width $R_0$ in the AS approximation is:

\be
R_{0AS}=\sqrt{\frac{8\pi }{P}} ,
\ee

\noindent and it is $\sqrt{2}$ times less than in the VA
approximation at a same power, figure \ref{Fig1}. The corresponding
stationary amplitudes in both approximations are presented in
figure \ref{Fig2}. Because of the relation $P=\pi A_0^2 R_0^2$,
the equilibrium amplitude $A_0$ in the AS approximation is $\sqrt{2}$
times greater than the one in the VA approximation at the same power.

The period of small oscillations of the
width perturbation around the equilibrium can be obtained from
(\ref{as4}):

\be
\Lambda _{AS}=\frac{8\sqrt{2}\pi^2}{P} .
\ee

\begin{figure}\vspace{0mm}
\includegraphics[width=70mm]{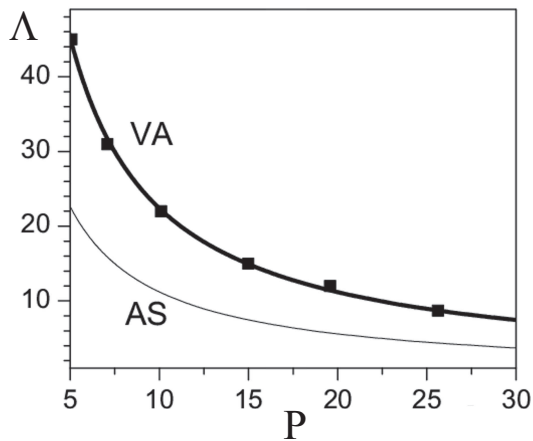}\vspace{0mm}
\caption{\label{Fig3}
Period of small oscillations as a function
of the beam power, in both approximations.
} \vspace{0mm}
\end{figure}

\noindent The period in the AS approximation is $2$ times less than
that in the VA approximation at the same power, figure \ref{Fig3}.
Thus, the values of the beam parameters for AS are systematically off
the values for the VA approximation, which, on the other hand,
happen to be very close to the full
numerical solution for the same values of parameters.

Another useful approximation to AS is based on the solution of
equation (\ref{eq2}) when the parameter $\theta
_2$ is independent of $z$.
In contrast to equation (\ref{as2}), in which $\theta_2$ may
depend on $z$, the equation
for $R$ has an exact oscillatory solution \cite{guo}

\be
R=R_0\sqrt{\cos ^2(\pi z/\Lambda )+\frac{P_0}{P}\sin^2(\pi z/\Lambda )} ,
\ee from which the solutions for $C$ and $\psi$ immediately follow:

\be
C=\frac{\pi }{\Lambda }\frac{\left( P_0/P-1\right) \sin (2\pi z/\Lambda )}{
4\left( \cos^2(\pi z/\Lambda )+\left( P_0/P\right) \sin
^2(\pi z/\Lambda )\right) } ,
\ee

\be
\psi = - \arctan \left( \sqrt{P_0/P}\tan (\pi z/\Lambda )\right) .
\ee

\noindent Here $R_0$ and $P_0$
are the width and the power of the AS solution, respectively. When $P=P_0$,
one obtains stationary AS; otherwise, the approximate solution
oscillates. The quantity
$\Lambda =\Lambda _{AS2}\sqrt{P_{0}/P}$ represents the period of harmonic
oscillations around the equilibrium (soliton) state, while
$\Lambda_{AS2}$ is the period of small oscillations of the width perturbation,

\be \Lambda _{AS2} \equiv \frac{8\sqrt{2}\pi ^2}{P}=\pi R_0^2 , \ee
which is the same as before. Thus, in the AS approximation one obtains
nice dependencies in closed form, but of little benefit,
in view of the large discrepancy with the VA and the numerical solution to the full problem.

\section{Conclusions}

In conclusion, we have discussed the differences between the VA and AS
approximate solutions to the propagation of solitons
in highly nonlocal nonlinear media. The AS model provides a radical
simplification and allows for an elegant description, but has a
limited practical relevance, mainly because of the competition
between the nonlocality  and the finite size of the sample. The VA solution is not
so simple, but works very well in the limited region of large nonlocality.
We have found that the AS approximation can differ up to two times, when compared
to the more realistic VA approximation and the numerical solution.\vspace{5mm}

\begin{acknowledgments}
This publication was made possible by NPRP Grants No.\# 09-462-1-074 and No.\# 5-674-1-114 from the Qatar National Research Fund (a member of the Qatar Foundation). The statements made herein are solely the responsibility of the authors. Work at the Institute of Physics Belgrade was supported by the Ministry of Science of the Republic of Serbia under the projects No. OI 171033 and No. 171006.
\end{acknowledgments}

\end{document}